\documentclass[aip,jcp,reprint,amsmath,amssymb,superscriptaddress,floatfix]{revtex4-2}

\usepackage[T1]{fontenc}
\usepackage{mathptmx}
\usepackage{amsmath,amssymb,bm,mathtools}
\usepackage{graphicx}
\usepackage{microtype}
\usepackage{xcolor}
\usepackage{etoolbox}
\usepackage[colorlinks=true,linkcolor=blue!65!black,citecolor=blue!65!black,urlcolor=blue!65!black]{hyperref}

\newcommand{\Tr}{\operatorname{Tr}}
\newcommand{\dd}{\mathrm{d}}
\newcommand{\ee}{\mathrm{e}}
\newcommand{\ii}{\mathrm{i}}
\newcommand{\hc}{\mathrm{h.c.}}
\newcommand{\ket}[1]{\lvert #1\rangle}
\newcommand{\bra}[1]{\langle #1\rvert}

\begin{document}

\title{Scalar Spin Chirality from Dissipative Pumping and Lamb-Shift Precession}
\author{YuanDong Wang}
\email[Corresponding author: ]{ydwang@cau.edu.cn}
\affiliation{Department of Applied Physics, College of Science, China Agricultural University, Beijing 100083, People's Republic of China}
\author{JianHua Wei}
\email[Corresponding author: ]{wjh@ruc.edu.cn}
\affiliation{Department of Physics, Renmin University of China, Beijing 100872, People's Republic of China}
\date{\today}

\begin{abstract}
How electrical driving generates and reverses scalar spin chirality in an interacting system with a real Hamiltonian and no magnetic flux remains a fundamental open question. We show that dissipative tunneling prepares an orbital pseudospin, while virtual charge fluctuations generate a noncollinear Lamb field that rotates it toward the chiral direction. Bias changes their relative orientation and thereby reverses the current. This reservoir-induced orbital Hanle effect converts a nonchiral orbital polarization into scalar spin chirality. Within the ground-state space, the circulating current  is determined by  the scalar spin chirality. Hierarchical-equations-of-motion calculations show that the low-bias reversal survives beyond the second-order weak-coupling description, while the quadratic growth of chirality after a sudden voltage switch identifies precession. More broadly, time-reversal-odd responses can emerge in open quantum systems from the interplay between dissipative state preparation and coherent precession.
\end{abstract}

\maketitle

\section{Introduction}

Scalar spin chirality,
\begin{equation}
\hat\chi=\bm S_1\cdot(\bm S_2\times\bm S_3),
\label{eq:chi}
\end{equation}
is the simplest spin-rotation-invariant observable that is odd under time
reversal. It characterizes the handedness of noncoplanar spin correlations and
provides a microscopic origin of Berry-phase Hall and thermal Hall responses
\cite{wen1989,kalmeyer1987,yang1993,ye1999,shindou2001,katsura2010}.
In equilibrium, scalar chirality is commonly associated with magnetic flux,
complex hopping processes, noncoplanar magnetic order, or spontaneous
time-reversal breaking
\cite{motrunich2006,bulaevskii2008,szasz2020,chen2022}.
The corresponding nonequilibrium problem is to generate and reverse this
handedness by electrical driving, without magnetic flux, complex tunneling
amplitudes, or an explicit chiral term in the microscopic Hamiltonian.

A triangular triple quantum dot (TQD) provides a minimal platform for addressing
this problem \cite{gaudreau2006,laird2010,acuna2024}.
In the three-electron Coulomb valley, its ground manifold contains a pair
of degenerate orbital states with opposite scalar chirality
\cite{scarola2004,trif2008,hsieh2012}.
These states also carry opposite circulating currents, so their orbital
magnetic moments provide a local probe of scalar spin chirality.
Circulating currents need not follow the lead current in an open ring
\cite{buttiker1983,bluhm2009,cini2010,lai2018}, allowing this internal
response to be distinguished from terminal transport.
Flux-induced chirality has been studied in triangular quantum dot molecules
\cite{garay2026}, while circulating currents and related control
schemes have been explored in quantum dot systems
\cite{wang2020,qi2023,qiwei2024,qi2024,dong2026}.

Electrical control in this setting requires understanding how biased
reservoirs prepare and evolve the degenerate orbital state.
Reservoir coupling is conventionally associated with level broadening and
decoherence \cite{datta1995,breuer2002}, but it can also generate orbital
coherence, while virtual charge fluctuations produce a matrix-valued
Lamb shift \cite{konig2003,braun2004,schultz2009,emary2007,donarini2010,
niklas2017,maurer2020}.
The Lamb shift can rotate the orbital polarization created by tunneling. Rate equations that track only orbital populations cannot capture this precession because they omit coherence between the degenerate states. 
A microscopic account is still needed of how these processes generate a
handed response under bias, what determines its reversal points, and how
the resulting scalar chirality is related to the circulating current.

\begin{figure}[!t]
\centering
\includegraphics[width=\columnwidth]{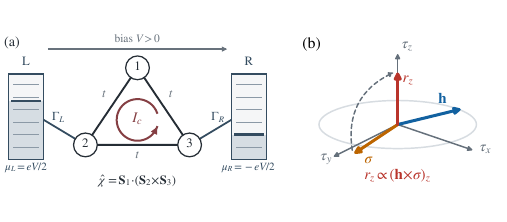}
\caption{Device and physical mechanism. (a) Dots 2 and 3 are coupled to
separate reservoirs, while dot 1 remains floating. The bias drives a circulating current $I_c$ without 
magnetic flux. (b) Orbital-pseudospin geometry of the dissipative source
$\bm\sigma$, projected Lamb field $\bm h$, and axial response $r_z$. The
stationary chiral response is controlled by
$(\bm h\times\bm\sigma)_z$.}
\label{fig:schematic}
\end{figure}

We resolve these questions by identifying a reservoir-induced orbital Hanle
effect \cite{hanle1924,jedema2002,zutic2004} in the chiral ground space of the triangular TQD. Sequential
tunneling prepares an in-plane orbital (chiral)
pseudospin, while virtual charge fluctuations generate a generally
noncollinear Lamb field that rotates it toward the chiral direction.  A coherence-retaining reduced dynamics separates the dissipative
source $\bm\sigma$ from the Lamb field $\bm h$ that produces the
precessional torque, and the  chirality is
controlled by $(\bm h\times\bm\sigma)_z$. Bias-driven reorientation of the two
vectors $\bm{h}$ and $\bm{\sigma}$ reverses the current whenever they become parallel or antiparallel, as illustrated in Fig.~\ref{fig:schematic}. Hierarchical-equations-of-motion 
(HEOM) benchmarks
\cite{tanimura1989,jin2008,hu2011,dan2023} show that the low-bias reversal
persists beyond the second-order weak-coupling description, while a sudden
voltage switch produces a characteristic quadratic short-time response.

\section{Model and circulating current}

We consider a triangular TQD with two of its dots coupled to
separate reservoirs, as shown in Fig.~\ref{fig:schematic}(a). Its Hamiltonian is
\begin{equation}
H_{\rm dot}=t\sum_{\langle ij\rangle s}
\left(c_{is}^{\dagger}c_{js}+\hc\right)
+U\sum_i n_{i\uparrow}n_{i\downarrow}+\epsilon_d N,
\label{eq:Hdot}
\end{equation}
where $t$ is the hopping, $U>0$ is the onsite repulsion,
$\epsilon_d$ is the gate energy, $n_{is}=c_{is}^{\dagger}c_{is}$, and
$N=\sum_{is}n_{is}$. Dots $i_L=2$ and $i_R=3$ couple to
normal reservoirs through
$H_T=\sum_{\ell kis}(V_{\ell k i} a_{\ell ks}^{\dagger}c_{i_\ell s}+\hc)$. The leads are represented by $H_{\rm res}=\sum_{\ell ks}\epsilon_{ \ell k}a_{\ell ks}^\dagger a_{\ell ks}$. 
Here $a_{\ell ks}$ annihilates a spin $s$ electron in reservoir $\ell$. 
The chemical potentials are $\mu_L=eV/2$ and $\mu_R=-eV/2$. The influence of the leads is fully captured by the hybridization functions $
J_{\ell i}(\omega)
=2\pi\sum_k|V_{\ell k i}|^2\delta(\omega-\epsilon_{\ell k})
$, which assume a Lorentz function $
J_{\ell i}(\omega)=
{\Gamma_{\ell i} W_{\ell}^2}/{((\omega - \mu_{\ell})^2+W_{\ell}^2)}
$. Here, $\Gamma_ {\ell i}$ and $W_{\ell}$ are the linewidth (hybridization strength) and the bandwidth of the leads, respectively. With dot 1
floating, the nonzero terms are $\Gamma_ {L 2}$ and $\Gamma_ {R 3}$. 
For simplicity
we assume $\Gamma_ {L 2} = \Gamma_ {R 3} =\Gamma$ and $W_L =W_R=W$. The gate is chosen in the three-electron Coulomb valley with $\epsilon_d = -U/2$.  

The circulating current is defined via the derivative of the 
Peierls phase (for details see Appendix~\ref{app:model}), which is
\begin{equation}
\hat I_c=-\frac{\ii et}{3\hbar}\sum_s
\left(c_{1s}^{\dagger}c_{2s}+c_{2s}^{\dagger}c_{3s}
+c_{3s}^{\dagger}c_{1s}-\hc\right).
\label{eq:Iop}
\end{equation}
The steady-state current pattern in Fig.~\ref{fig:schematic}(a) follows from
charge conservation. Because no charge accumulates on the floating dot, the
bond currents decompose into source-to-drain transport and a divergence-free
circulating component of equal magnitude on all three bonds. The latter is
measured by Eq.~\eqref{eq:Iop}.

In the Coulomb valley of interest, the isolated TQD has a fourfold ground
multiplet with three electrons. Let $P$ project onto this space. It can be proved that the 
circulating current within the ground multiplet is (see Appendix~\ref{app:model})
\begin{equation}
\hat{I}_{c}^{(P)} = P\hat I_cP=\lambda P\hat\chi P
=\lambda\chi_q\hat\tau_z,
\label{eq:projection}
\end{equation}
where $P\hat\chi P=\chi_q\hat\tau_z$ defines the chirality pseudospin and
$\hat{\bm\tau}=(\hat\tau_x,\hat\tau_y,\hat\tau_z)$ denotes Pauli operators
on the chiral (orbital) doublet and the identity on physical spin, normalized by
$\Tr(P)=4$ and
$\Tr(\hat\tau_a\hat\tau_b)=4\delta_{ab}$. The coefficient
$\lambda$ is fixed by the isolated dot. In the strong-coupling limit
$t/U\to0$,
$\chi_q\to\sqrt3/4$ and
$\lambda\to-24et^3/(\hbar U^2)$. 
Although the current--chirality relation is usually derived in the large-$U$ limit, Eq.~\eqref{eq:projection} is exact within the ground multiplet at finite $t/U$.  In the spin-only
limit, the two orbital states are the eigenstates of cyclic permutation and
have chirality eigenvalues $\pm\sqrt3/4$. At finite $U$, charge fluctuations
renormalize both $\chi_q$ and $\lambda$, while their product remains the
matrix element of the current operator.

\section{Reduced dynamics and chiral response}

Let $\varrho=\Tr_{\rm res}\varrho_{\rm tot}$ denote the full dot RDM on the
complete dot Fock space, and let $Q_2$ and $Q_4$ project onto the lowest two-
and four-electron multiplets, respectively. These charged multiplets carry
one-dimensional orbital representations and are a spin singlet and a spin
triplet, respectively. In the transport window considered here,
sequential-tunneling transitions between $P$ and higher charged multiplets
are off shell. We therefore restrict the $P$-adjacent jump operators to
$Q_2$ and $Q_4$, while retaining all real transitions that do not involve
$P$. This restriction applies only to the dissipative kernel and the Lamb shift
retains the complete two- and four-electron virtual spectrum. With all rates
expressed in energy units, the Born, Markov, and secular equation is
\begin{equation}
\dot\varrho=-\frac{\ii}{\hbar}[H_{\rm dot}+H_{\rm LS},\varrho]
+\frac{1}{\hbar}\mathcal D_{\rm seq}(\varrho).
\label{eq:master}
\end{equation}
The sequential-tunneling kernel has an explicit
Gorini--Kossakowski--Sudarshan--Lindblad (GKSL) form on the full Fock space and
retains all coherences inside the  degenerate ground space.
Treating the ground multiplet in terms of populations alone would discard its
orbital coherences and produce basis-dependent results. The complete
construction is given in Appendix~\ref{app:master}. We refer to Eq.~\eqref{eq:master}, with the restricted
$P$-adjacent channels specified above, as the channel-restricted second-order
GKSL equation. Projecting only the state onto $P$, we write
\begin{equation}
\rho_P\equiv P\varrho P
=\frac{w}{4}\left(P+\bm r\cdot\hat{\bm\tau}\right),
\qquad
PH_{\rm LS}P=h_0P+\bm h\cdot\hat{\bm\tau},
\label{eq:definitions}
\end{equation}
where $w=\Tr\rho_P$ is the ground-multiplet probability,
$\bm r=(r_x,r_y,r_z)$ is its conditional chirality pseudospin, $h_0$ is a
scalar shift, and $\bm h=(h_x,h_y,h_z)$ is the projected Lamb field. We define
the ground-multiplet contribution to the current by
$I_c^{(P)}\equiv\Tr(\rho_P\hat I_c^{(P)})$ and the complementary contribution
by $I_c^{(\mathrm{rest})}\equiv I_c-I_c^{(P)}$. The latter collects the
retained charged and excited three-electron sectors. It is small in the
central Coulomb valley and becomes relevant near a charge threshold.
Because $P$ is an exactly degenerate eigenspace,
$PH_{\rm dot}P=E_0P$ and its commutator with $\rho_P$ vanishes. In contrast,
the Pauli algebra $[\hat\tau_a,\hat\tau_b]=2\ii\epsilon_{abc}\hat\tau_c$
converts the non-scalar Lamb shift into
\begin{equation}
-\frac{\ii}{\hbar}\Tr_P\!\left[
\hat{\bm\tau}[PH_{\rm LS}P,\rho_P]\right]
=\frac{2w}{\hbar}\,\bm h\times\bm r.
\end{equation}
The cross product therefore describes Lamb-shift-induced precession within
the degenerate chiral doublet. Including dissipative pumping yields a Bloch equation for the orbital pseudospin
\begin{equation}
\frac{\dd(w\bm r)}{\dd t}=\frac{w}{\hbar}
\left[2\bm h\times\bm r+\bm\sigma(t)-\gamma\bm r\right].
\label{eq:bloch}
\end{equation}
Here $\bm\sigma(t)$ is the effective pumping source and $\gamma$ is
the isotropic loss coefficient, both obtained by projecting the
sequential-tunneling terms of the microscopic master equation onto
the ground multiplet.   Equation~\eqref{eq:bloch} is the central result of this study. It reveals the dynamics of the emergence of the chirality. That is, $\bm\sigma$ pumps an in-plane polarization,
$\bm h$ induces precession, and $\gamma$ relaxes the pseudospin. For the tunneling transitions retained in our zero-flux model,
the source $\bm\sigma(t)$ and the Lamb field $\bm h$ satisfy
\begin{equation}
\sigma_z(t)=h_z=0.
\end{equation}
Both vectors therefore lie in the pseudospin $xy$ plane.
Tunneling can build an in-plane orbital polarization, but it
does not directly pump the chiral component $r_z$.
A chiral response arises when the precession term
$2\bm h\times\bm r$ rotates this polarization out of the plane. The microscopic definitions and the derivation of the pseudospin equation
are given in Appendix~\ref{app:bloch}.

\begin{figure}[!t]
\centering
\includegraphics[width=\columnwidth]{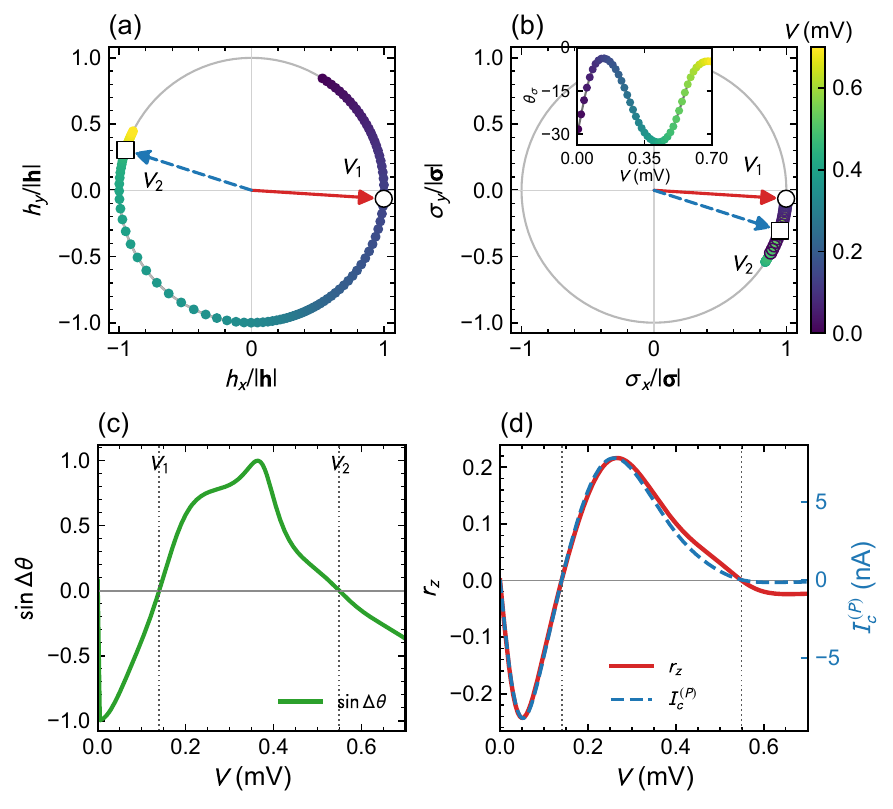}
\caption{Circulating current reversals governed by the relative orientation of
 the Lamb field and dissipative pump.
(a) Bias versus  Lamb-field direction $\bm h/|\bm h|$.
(b) Bias versus  pump direction $\bm\sigma/|\bm\sigma|$.
 The inset shows its polar angle $\theta_\sigma$.
The marked voltages $V_1$ and $V_2$ identify parallel and
antiparallel orientations of the two vectors, respectively.
(c) The angular factor $\sin\Delta\theta$, with
$\Delta\theta=\theta_\sigma-\theta_h$ and $\theta_h$ the polar angle
of the Lamb field.
(d) The axial pseudospin component $r_z$ and its ground-multiplet
projected current $I_c^{(P)}$. Parameters are $t=0.25$, $U=1$, $\epsilon_d=-\epsilon_0=-0.5$, $\Gamma_L=\Gamma_R=0.05$, $k_BT=0.05$, and $W=2$, in meV. }
\label{fig:geometry}
\end{figure}

At stationarity, 
\begin{equation}
0=2\bm h\times\bm r
+\bm\sigma-\gamma\bm r.
\label{eq:stationarybloch}
\end{equation}
Solving Eq.~\eqref{eq:stationarybloch} with $h_z=\sigma_z=0$ gives
\begin{equation}
r_z=\frac{2(\bm h\times\bm\sigma)_z}
{\gamma^2+4|\bm h|^2},
\label{eq:central}
\end{equation}
as illustrated in Fig.~\ref{fig:schematic}(b). 
Combining Eqs.~\eqref{eq:projection} and \eqref{eq:definitions} with the
strong-coupling coefficients stated above, the stationary ground-multiplet
current is
\begin{equation}
I_c^{(P)}=
-\frac{6\sqrt3\,e t^3}{\hbar U^2}\,w r_z,
\qquad t\ll U.
\end{equation}
The exact finite-$t/U$ projected-current relation, together with its
construction  is given in Appendix~\ref{app:model}.
A finite circulating current within the ground multiplet requires a nonzero population, orbital coherence, and a dissipative pump that is not collinear with the Lamb field.

Equation~\eqref{eq:bloch} therefore establishes the chiral response as a
reservoir-induced orbital Hanle effect. Unlike the
conventional Hanle setting, however, the effective field is generated by the
same reservoirs that create and relax the pseudospin. It therefore changes
with voltage, and gate position. The ratio
$2|\bm h|/\gamma$ separates overdamped and
precessional regimes. The competition between precession and relaxation is explicit upon writing
\begin{equation}
r_z=\frac{|\bm\sigma|}{\gamma}\sin\theta\,
\frac{x}{1+x^2},
\qquad
x=\frac{2|\bm h|}{\gamma},
\end{equation}
where $\theta$ is the angle from $\bm h$ to $\bm\sigma$.
For $x\ll1$, the pumped polarization decays before precession can
rotate it appreciably toward the chiral direction, giving
$r_z\simeq2(\bm h\times\bm\sigma)_z/\gamma^2$.
For $x\gg1$, rapid precession produces contributions of opposite
axial sign that partially cancel in the stationary state. In addition, 
although pumping, precession, and damping are all second order in the
tunnel amplitudes, their common linewidth dependence cancels from
$r_z$. Weak reservoir coupling can therefore induce a finite chirality. 

\section{Bias dependence and current reversals}

The bias dependence of the Lamb field originates from virtual
electron-removal and addition processes through the two- and
four-electron multiplets.
For each reservoir $\ell$, we define its Lamb field by projecting
its contribution $H_{\rm LS}^{\ell}(\mu_\ell)$ onto the ground multiplet:
\begin{equation}
PH_{\rm LS}^{\ell}(\mu_\ell)P
=h_{0,\ell}P+\bm h_\ell(\mu_\ell)\cdot\hat{\bm\tau}.
\end{equation}
Here $h_{0,\ell}$ shifts both orbital states equally, while
$\bm h_\ell$ governs their pseudospin precession.
The microscopic expression for $H_{\rm LS}^{\ell}$ and its
principal-value coefficients are given in Appendix~\ref{app:master}. 
For independent reservoirs, the second-order Lamb shift is additive,
$H_{\rm LS}=H_{\rm LS}^{L}+H_{\rm LS}^{R}$.
Adding the projected contributions and comparing with
Eq.~\eqref{eq:definitions} therefore gives
$\bm h=\bm h_L(\mu_L)+\bm h_R(\mu_R)$.
With $\mu_{L,R}=\pm eV/2$, this becomes
\begin{equation}
\bm h(V)
=\bm h_L(eV/2)+\bm h_R(-eV/2).
\label{eq:lambfield}
\end{equation}
Because the leads are attached to different dots, their Lamb fields
act along different pseudospin directions.
Changing the bias modifies the relative strengths of these
contributions. Their vector sum can therefore rotate with bias,
as shown by the trajectory of $\bm h/|\bm h|$ in
Fig.~\ref{fig:geometry}(a). 
Figure~\ref{fig:geometry}(b) shows the direction
$\bm\sigma/|\bm\sigma|$, with its  polar angle
$\theta_\sigma(V)=\operatorname{atan2}\!\left[\sigma_{y}(V),\sigma_{x}(V)\right]$ shown  in the inset.
It is seen that $\bm{\sigma}$ varies over a smaller angular range than $\bm{h}$.

According to Eq.~\eqref{eq:central}, $r_z$ vanishes when
$\bm h$ and $\bm\sigma$ are parallel or antiparallel:
\begin{equation}
h_x\sigma_y-h_y\sigma_x=0.
\label{eq:reversal}
\end{equation}
The marked directions in Figs.~\ref{fig:geometry}(a) and
\ref{fig:geometry}(b) are parallel at $V_1$ and antiparallel at $V_2$.
Figure~\ref{fig:geometry}(c) shows the corresponding angular factor
$\sin\Delta\theta$, where $\Delta\theta=\theta_\sigma-\theta_h$.
The angular factor changes sign at both alignments, reversing
$(\bm h\times\bm\sigma)_z$ and hence $r_z$. 
Figure~\ref{fig:geometry}(d) shows the resulting reversals of
$r_z$ and $I_c^{(P)}$ at the same voltages $V_1$ and $V_2$.
The denominator of Eq.~\eqref{eq:central} is strictly positive,
so these zero crossings are set by the relative orientation in
Figs.~\ref{fig:geometry}(a)--\ref{fig:geometry}(c).

Mirror covariance under contact exchange gives
$h_x(-V)=h_x(V)$ and $h_y(-V)=-h_y(V)$, with
the corresponding parity for the stationary pump. Consequently,
$h_y(0)=\sigma_y(0)=0$. Together with $h_z=\sigma_z=0$, this makes $\bm{h}$ and $\bm{\sigma}$ collinear along the $x$ direction at zero bias. Equation
\eqref{eq:central} then gives $r_z(0)=0$ even when $h_x(0)$ and
$\sigma_x(0)$ remain finite. The componentwise derivation is given in Appendix~\ref{app:bloch}. It also follows that
$I_c^{(P)}(-V)=-I_c^{(P)}(V)$, as required for a bias-generated handedness.
Figure~\ref{fig:app:gate} in Appendix~\ref{app:gate} shows that the alignment condition tracks the total-current $I_c=0$
throughout the ground-multiplet regime, while the high-bias separation between
$I_c=0$ and $I_{c}^{(P)}=0$ marks the onset of a sizable
$I_c^{(\mathrm{rest})}$, rather than a failure of Eq.~\eqref{eq:central}.

\section{Formation and relaxation of scalar spin chirality}
\subsection{Linewidth scaling and comparison with HEOM}

To assess the predictions of the second-order master equation,
we first establish how the circulating and transport currents
scale with the reservoir linewidths, then compare these scalings
and the current reversals with HEOM calculations.
Let $\Gamma_\ell=\eta\bar\Gamma_\ell$ with $\eta>0$, keeping the
 other parameters fixed.
The sequential-tunneling kernel and Lamb shift are both linear
in the linewidths:
${\cal D}_{{\rm seq},\eta}=\eta\bar{\cal D}_{\rm seq}$ and
$H_{{\rm LS},\eta}=\eta\bar H_{\rm LS}$.
In the energy-block-diagonal sector, where
$[H_{\rm dot},\varrho]=0$, the secular equation therefore gives
\begin{equation}
\dot\varrho_\eta=\eta\bar{\cal L}_{\rm seq}\varrho_\eta,
\label{eq:scaledgenerator}
\end{equation}
where
\begin{equation}
\bar{\cal L}_{\rm seq}X=\frac1{\hbar}
[-\ii[\bar H_{\rm LS},X]+\bar{\cal D}_{\rm seq}(X)].
\end{equation}
For a unique normalized stationary state in this sector,
$\varrho_{{\rm ss},\eta}=\bar\varrho_{\rm ss}$. 
The circulating-current operator is independent of the reservoir
linewidths, whereas the lead current is linear in the tunneling rates.
Consequently,
\begin{equation}
I_{c,\eta}^{\rm seq}
=\Tr(\bar\varrho_{\rm ss}\hat I_c)
=\bar I_c^{\rm seq}.
\label{eq:currentscaling}
\end{equation}
The current from lead $\ell$ instead obeys
\begin{equation}
I_{\ell,\eta}
=\frac{e}{\hbar}
\Tr\!\left[N{\cal D}_{\ell,\eta}
(\bar\varrho_{\rm ss})\right]
=\eta\bar I_\ell.
\end{equation}
Here ${\cal D}_{\ell,\eta}$ is the contribution of lead $\ell$ to the
sequential-tunneling kernel, and $I_\ell>0$ denotes electron flow
from that lead into the dot.
At stationarity, $I_L+I_R=0$, so the transport current
$I_{\rm tr}=(I_L-I_R)/2$ also scales linearly with $\eta$.

The same scaling follows from the pseudospin description.
Writing $\bm h_\eta=\eta\bar{\bm h}$,
$\bm\sigma_{\eta}=\eta\bar{\bm\sigma}$,
and $\gamma_\eta=\eta\bar\gamma$, Eq.~\eqref{eq:central} gives
\begin{equation}
r_{z,\eta}
=\frac{2\eta^2(\bar{\bm h}\times\bar{\bm\sigma})_z}
{\eta^2(\bar\gamma^2+4|\bar{\bm h}|^2)}
=\bar r_z.
\end{equation}
Together with $w_\eta=\bar w$, which follows from the
linewidth-independent stationary state, this yields
\begin{equation}
I_{c,\eta}^{(P)}=\lambda\chi_q\bar w\,\bar r_z
=\bar I_c^{(P)}.
\end{equation}
The transport-current scaling is
\begin{equation}
I_{{\rm tr},\eta}=\eta\bar I_{\rm tr}.\label{Itr-scale}
\end{equation}
These scaling relations hold for the specified second-order
secular equation under a common rescaling of the  linewidth.

\begin{figure}[!t]
\centering
\includegraphics[width=\columnwidth]{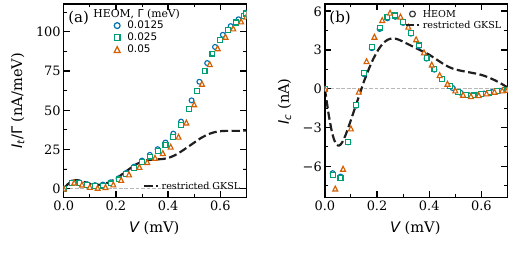}
\caption{Weak-coupling benchmark against HEOM.
Open symbols show results for three values of $\Gamma$. Black dashed curves
show the channel-restricted second-order GKSL result. (a)  Transport current (divided by $\Gamma$) versus bias $V$ for different $\Gamma$.  (b) Circulating current versus bias $V$ for different $\Gamma$. The other 
parameters are the same as those in  Fig.~\ref{fig:geometry}.}
\label{fig:benchmark}
\end{figure}

In Fig~\ref{fig:benchmark}(a) we compare the transport current from
HEOM with the GKSL result. 
The HEOM results for $I_{\rm tr}/\Gamma$ are nearly independent
of the linewidth, consistent with Eq.~(\ref{Itr-scale}). 
The GKSL result agrees with HEOM at low bias but
underestimates the transport current at higher bias, where
additional sequential-tunneling transitions become relevant.
Including these transitions improves the agreement, as shown
in Appendix~\ref{app:comparison}. 
Figure~\ref{fig:benchmark}(b) shows qualitative agreement between
the GKSL and HEOM results. They both predict a nonmonotonic bias
dependence and a low-bias reversal of the circulating current.
The HEOM currents at the three linewidths are also close in
magnitude without normalization by $\Gamma$, in contrast to
the transport current in Fig.~\ref{fig:benchmark}(a).
This weak linewidth dependence supports the prediction that
the leading stationary circulating current is independent
of $\Gamma$ and remains finite in the weak-coupling limit.

\subsection{Transient response to a bias switch}

\begin{figure}[!t]
\centering
\includegraphics[width=\columnwidth]{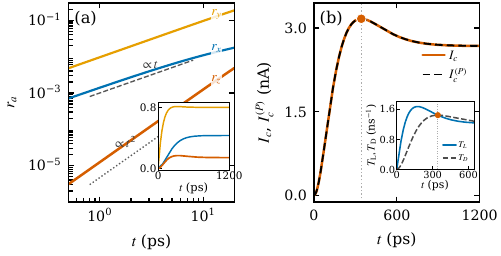}
\caption{Pseudospin and circulating-current dynamics following a bias quench.
(a) Dissipative pumping initially drives a linear growth of the in-plane components $r_x$ and $r_y$, while Lamb-induced precession generates a quadratic onset of $r_z$. The inset shows the subsequent relaxation toward the steady state.
(b) Total circulating current $I_c$ and its ground-multiplet contribution $I_c^{(P)}$. The inset compares the Lamb torque and damping. The parameters are the same as those in  Fig.~\ref{fig:geometry}.}
\label{fig:dynamics}
\end{figure}

To understand how the bias-induced chirality develops and on
what time scale, we follow the circulating current after a
sudden switch from equilibrium to finite bias. 
Define
$\bm u=w\bm r$.
The axial component of Eq.~\eqref{eq:bloch} gives
\begin{equation}
\hbar\dot u_z=2(\bm h_f\times\bm u)_z-\gamma u_z,
\label{eq:quenchaxial}
\end{equation}
where the post-quench Lamb field $\bm h_f$ and relaxation rate
$\gamma$ are constant. The $\epsilon_d = -U/2$ equilibrium initial state is unpolarized with $\rho_{P,i}=w_iP/4$ and  $w_i=w(0)$.
Hence $\bm u_i=\bm u(0)=0$.
Defining $\bm S_i=w_i\bm\sigma(0^+)$,  
the in-plane components of Eq.~\eqref{eq:bloch} then give
\begin{equation}
\hbar\dot u_a(0^+)=S_{i,a},\qquad a=x,y.
\label{eq:initialinplane}
\end{equation}
Thus $u_a(t)=S_{i,a}t/\hbar+O(t^2)$.
Using $w(t)=w_i+O(t)$, we obtain
\begin{equation}
r_a(t)=\frac{\sigma_a(0^+)}{\hbar}t+O(t^2),\qquad a=x,y.
\label{eq:inplaneshorttime}
\end{equation}
Both the in-plane components therefore grow linearly, as shown in
Fig.~\ref{fig:dynamics}(a). 
Since $\bm u_i=0$, Eq.~\eqref{eq:quenchaxial} gives
$\dot u_z(0^+)=0$.
Differentiating it and using Eq.~\eqref{eq:initialinplane} yields
\begin{equation}
\ddot u_z(0^+)=\frac{2}{\hbar^2}(\bm h_f\times\bm S_i)_z,
\label{eq:initialacceleration}
\end{equation}
so the short-time expansion is
\begin{equation}
u_z(t)=\frac{(\bm h_f\times\bm S_i)_z}{\hbar^2}t^2+O(t^3).
\label{eq:shorttime}
\end{equation}
Consequently, both $r_z=u_z/w$ and
$I_c^{(P)}=\lambda\chi_q u_z$ are quadratically dependent on $t^2$. The inset of Fig.~\ref{fig:dynamics}(a) shows how the components
depart from these short-time power laws and approach their
stationary values at longer times.

Figure~\ref{fig:dynamics}(b) shows the circulating current
rising to a maximum before relaxing to its stationary value.
The near coincidence of $I_c^{(P)}$ and $I_c$ indicates that
the ground multiplet dominates this evolution.
The inset explains the current maximum through the balance
between the Lamb torque and damping:
\begin{equation}
\frac{\dd I_c^{(P)}}{\dd t}
=\lambda\chi_q(T_{\rm L}-T_{\rm D}),
\end{equation}
where $T_{\rm L}=2w(\bm h_f\times\bm r)_z/\hbar$ and
$T_{\rm D}=w\gamma r_z/\hbar$.
Their crossings change the sign of the current derivative
and therefore mark extrema of $I_c^{(P)}$.
The peak thus marks the balance between the buildup of chirality
through Lamb-field precession and its decay through dissipative
relaxation.

\subsection{Experimental detection of the circulating current}

The circulating current provides a magnetic signature of the
chiral state through an orbital moment normal to the triangular
plaquette. For an equilateral device with side length $a=200$ nm,
the enclosed area is $A=\sqrt3a^2/4$. A current of $I_c=3$ nA
then produces a magnetic moment
$I_cA=5.2\times10^{-23}\,{\rm A\,m^2}\simeq5.6\,\mu_B$.
A Biot--Savart estimate gives $|B_z|\simeq14$ nT at a point
$50$ nm above the center of the plaquette. This estimate suggests
nanoscale SQUID probes as a possible means of detecting the
stationary circulating current \cite{vasyukov2013}.

Resolving the transient requires sufficient temporal resolution
as well as magnetic sensitivity. The current maximum at $343$ ps
sets a subnanosecond time scale, corresponding to a characteristic
frequency of about $2.9$ GHz. While a scanning probe could measure
the stationary field, the transient could be accessed through
repeated pump--probe sampling or an on-chip resonator.
Recent quantum magnetometry has demonstrated a time resolution
of $1.1$ ns \cite{herb2025}, approaching the subnanosecond
resolution needed to follow the predicted current peak.

\section{Conclusions}

In summary, a dc bias generates and reverses scalar spin chirality through
reservoir-induced  precession of an orbital pseudospin. Within the Coulomb valley regime, the circulating-current  is  proportional to the scalar spin chirality, which is determined by the relative orientation of the dissipative pump
and Lamb field. HEOM calculations show that the
low-bias reversal persists beyond the second-order weak-coupling description,
and the quadratic short-time response provides a
dynamical signature of the torque mechanism.

Our results show that a time-reversal-odd
response can emerge even when the microscopic Hamiltonian
preserves time-reversal symmetry and tunneling provides
no direct source of chirality. Tunneling generates an in-plane orbital polarization, which
the reservoir-induced Lamb field converts into chirality
through coherent precession. Similar torques may act on plaquette, sublattice, or
momentum-space pseudospins in extended systems
\cite{diehl2008,yanay2018,dogra2019,szasz2020,chen2022}. Nonequilibrium cluster
methods could determine whether such local torques can stabilize or switch
bulk chiral order \cite{maier2005,aoki2014,arrigoni2013}. Specifically,
nonequilibrium cellular DMFT would treat a triangular plaquette as an impurity
coupled self-consistently to the surrounding lattice, allowing the  mechanism we discovered  to be tested beyond an isolated cluster.
Calculating the lattice-averaged scalar chirality and its stability
under bias would show whether the local torque can sustain bulk
chiral order when feedback from the surrounding lattice is included. Reservoir-induced virtual processes could therefore serve as
an electrically tunable ordering field for macroscopic time-reversal breaking,
enabling the selection and switching of chiral phases without an explicit
chiral term in the microscopic Hamiltonian.

\begin{acknowledgments}
We thank Chen Peng for inspiring discussions. This work is supported by the
National Natural Science Foundation of China through Grant No.~12404147.
\end{acknowledgments}

% These statements require information supplied by the authors.
% Do not replace the placeholders with factual claims without confirmation.
\section*{Author declarations}
\subsection*{Conflict of interest}
The authors have no conflicts to disclose.

\subsection*{Author Contributions}\noindent\textbf{YuanDong Wang}: Conceptualization (lead); Writing – original draft (equal); Writing – review $\&$ editing (equal). \textbf{JianHua Wei}: Conceptualization (supporting); Writing – original draft (equal); Writing – review $\&$ editing (equal).

\appendix
% Appendices are kept inline in this main document.

\section{Current--chirality relation in the ground multiplet}
\label{app:model}

A flux $\Phi$ through the triangle introduces the Peierls factor
$\exp(-\ii\theta/3)$ on each directed hopping term in Eq.~\eqref{eq:Hdot},
where $\theta=2\pi\Phi/\Phi_0=e\Phi/\hbar$. The circulating current is
\begin{equation}
\hat I_c=-\left.\partial_\Phi H_{\rm dot}(\Phi)\right|_0
=+\frac{\ii et}{3\hbar}\sum_s
(c_{1s}^{\dagger}c_{2s}+c_{2s}^{\dagger}c_{3s}
+c_{3s}^{\dagger}c_{1s}-\hc).
\label{eq:app:Iop}
\end{equation}
Equivalently, the bond particle current in energy units is
$\hat J_{i\to j}=\ii t\sum_s
(c_{is}^{\dagger}c_{js}-c_{js}^{\dagger}c_{is})$.
The isolated-dot continuity equation,
$\hbar\dot n_i=-\sum_j\hat J_{i\to j}$, gives
\begin{equation}
\hat I_c=\frac{e}{3\hbar}
(\hat J_{1\to2}+\hat J_{2\to3}+\hat J_{3\to1}).
\label{eq:app:Iaverage}
\end{equation}
Thus the circulating current averages the three bond currents, with positive
circulation along $1\to2\to3\to1$. Reversing this orientation reverses
both $\hat I_c$ and $\lambda$. In a driven state, the flux derivative of
the energy expectation generally includes the flux dependence of the
density operator and need not equal Eq.~\eqref{eq:app:Iop}
\cite{byers1961,cini2010}.

In the three-electron Coulomb valley, $P$ projects onto the fourfold ground
multiplet: a spin doublet and an orbital doublet. For the present parameters
($t,U>0$), the lowest two- and four-electron multiplets are an
orbital $A_1$ spin singlet and an orbital $A_2$ spin triplet, respectively.
Their projectors are
\begin{equation}
\begin{aligned}
Q_2&=|\psi_{2,0}\rangle\langle\psi_{2,0}|,\\
Q_4&=\sum_{M=-1}^{1}|\psi_{4,0;M}\rangle\langle\psi_{4,0;M}|.
\end{aligned}
\label{eq:app:chargedprojectors}
\end{equation}
The states are orthonormal ground eigenstates of $H_{\rm dot}$ in their
respective charge sectors, with $M$ denoting the total-spin $z$ component.
Thus $\Tr Q_2=1$ and $\Tr Q_4=3$. We write $Q_{\rm ch}=Q_2+Q_4$;
these projectors are real at zero flux.

The scalar spin chirality is defined as
\begin{equation}
\hat\chi=\bm S_1\cdot(\bm S_2\times\bm S_3),\qquad
\bm S_i=\frac12\sum_{ss'}c_{is}^{\dagger}\bm\sigma_{ss'}c_{is'}.
\label{eq:app:chi}
\end{equation}
In the singly occupied $S^z_{\rm tot}=1/2$ sector, take
$\ket1=\ket{\downarrow\uparrow\uparrow}$,
$\ket2=\ket{\uparrow\downarrow\uparrow}$, and
$\ket3=\ket{\uparrow\uparrow\downarrow}$. With the cyclic permutation
$P_{123}\ket j=\ket{j+1}$ (indices modulo three),
\begin{equation}
\hat\chi=\frac{\ii}{4}(P_{123}-P_{132}).
\end{equation}
Its chiral eigenstates are
\begin{equation}
\ket{\chi_\pm}=\frac{\ket1+\omega_\pm\ket2+\omega_\pm^2\ket3}{\sqrt3},
\label{eq:app:chistates}
\end{equation}
where $\omega_\pm=\ee^{\pm2\pi\ii/3}$. These states satisfy
\begin{equation}
\hat\chi\ket{\chi_\pm}=\pm\frac{\sqrt3}{4}\ket{\chi_\pm}.
\end{equation}
Define $\hat\tau_z=\ket{\chi_+}\bra{\chi_+}-\ket{\chi_-}\bra{\chi_-}$,
$\hat\tau_x=\ket{\chi_+}\bra{\chi_-}+\ket{\chi_-}\bra{\chi_+}$, and
$\hat\tau_y=\ii\hat\tau_x\hat\tau_z$, tensored with the identity on
physical spin. At finite $U$, we use the chiral basis of the exact ground
multiplet, with $\Tr P=4$ and $\Tr(\hat\tau_a\hat\tau_b)=4\delta_{ab}$.
Complex conjugation $K$ in the real occupation basis and the reflection
$\mathcal M$ exchanging sites 2 and 3 satisfy
\begin{equation}
 K\hat\tau_{x,y} K=\hat\tau_{x,y},\qquad
 K\hat\tau_z K=-\hat\tau_z,\qquad
P\mathcal M P=\hat\tau_x.
\label{eq:app:Ktau}
\end{equation}
Since $\hat I_c$ and $\hat\chi$ are spin scalars and odd under $K$,
their projections onto the ground multiplet are proportional to $\hat\tau_z$:
\begin{equation}
P\hat\chi P=\chi_q\hat\tau_z,
\end{equation}
and hence
\begin{equation}
P\hat I_cP=\lambda\chi_q\hat\tau_z=\lambda P\hat\chi P.
\label{eq:app:exact}
\end{equation}
The coefficients are given by
\begin{equation}
\chi_q=\frac14\Tr(\hat\tau_zP\hat\chi P)
\end{equation}
and
\begin{equation}
\lambda=\frac{\Tr(\hat\tau_zP\hat I_cP)}{4\chi_q}.
\end{equation}
Equation~\eqref{eq:app:exact} includes charge fluctuations at finite $U$.
Its strong-coupling limit follows by eliminating double occupancy. Let
$P_0$ project onto single occupancy, $T_\theta$ denote the
hopping operator, and $R=Q_0(E_{\rm so}-H_U)^{-1}Q_0$, with $Q_0=1-P_0$
in the three-electron sector. Here $H_U$ is the on-site interaction;
subtracting the common gate energy gives $E_{\rm so}=0$. Each three-hop
process around the triangle has two intermediate states at excitation
energy $U$, giving the denominator $U^2$. Summing the ordered paths yields
the third-order term odd in flux,
\begin{equation}
\begin{aligned}
H_{\rm eff,odd}^{(3)}
&=\tfrac12\bigl(P_0T_\theta RT_\theta RT_\theta P_0\\
&\qquad-P_0T_{-\theta}RT_{-\theta}RT_{-\theta}P_0\bigr)\\
&=\frac{6\ii t^3}{U^2}\sin\theta(P_{123}-P_{132})\\
&=\frac{24t^3}{U^2}\sin\theta\,\hat\chi.
\end{aligned}
\label{eq:app:strong}
\end{equation}
Flux differentiation at zero flux gives $\lambda\to-24et^3/(\hbar U^2)$
for the orientation in Eq.~\eqref{eq:app:Iop}, with $\chi_q\to\sqrt3/4$
in the singly occupied limit.

For the full dot density operator $\varrho$, we write
\begin{equation}
I_c^{(P)}=\Tr(P\varrho P\hat I_c),\qquad
I_c=I_c^{(P)}+I_c^{(\rm rest)}.
\label{eq:app:currentdecomp}
\end{equation}
The remainder $I_c^{(\rm rest)}$ includes charged and excited three-electron
states, for which Eq.~\eqref{eq:app:exact} does not imply a proportionality
between current and chirality.

\section{Degeneracy-preserving master equation}
\label{app:master}

The normalized dot density operator is
$\varrho=\Tr_{\rm res}\varrho_{\rm tot}$. The tunneling Hamiltonian is
$H_T=\sum_{\ell s}(d_{\ell s}\otimes B_{\ell s}^{\dagger}
+d_{\ell s}^{\dagger}\otimes B_{\ell s})$, where
$d_{\ell s}=c_{i_\ell s}$ and $B_{\ell s}=\sum_kV_\ell^*a_{\ell ks}$.
For initially stationary Gaussian reservoirs, the Born--Markov equation
in the interaction picture reads \cite{breuer2002}
\begin{equation}
\dot\varrho_I=-\frac1{\hbar^2}\int_0^\infty\dd u\,
\Tr_{\rm res}[H_{T,I}(t),[H_{T,I}(t-u),
\varrho_I(t)\otimes\rho_{\rm res}]].
\label{eq:app:BornMarkov}
\end{equation}
Let $\Pi_E$ project onto the complete eigenspace of $H_{\rm dot}$ at
energy $E$. The corresponding transition operators are
\begin{equation}
J_+^{\ell s}(\omega)=\sum_{E-E'=\omega}\Pi_Ed_{\ell s}\Pi_{E'},
\label{eq:app:eigenoperators}
\end{equation}
\begin{equation}
J_-^{\ell s}(\omega)=\sum_{E-E'=\omega}\Pi_Ed_{\ell s}^{\dagger}\Pi_{E'}.
\label{eq:app:eigenoperatorsminus}
\end{equation}
The secular approximation removes terms coupling distinct Bohr frequencies
but retains all coherences within each degenerate eigenspace.
The real parts of the reservoir integrals give the tunneling rates
\begin{equation}
\gamma_+^\ell(\omega)=\Gamma[1-f_{\mu_\ell}(-\omega)]\Theta(W-|\omega|),
\label{eq:app:rates}
\end{equation}
\begin{equation}
\gamma_-^\ell(\omega)=\Gamma f_{\mu_\ell}(\omega)\Theta(W-|\omega|).
\label{eq:app:ratesminus}
\end{equation}
The Lamb-shift coefficients are given by the principal-value integrals
\begin{equation}
s_+^\ell(\omega)=-\frac{\Gamma}{2\pi}{\cal P}\int_{-W}^{W}\dd\epsilon\,
\frac{1-f_{\mu_\ell}(\epsilon)}{\omega+\epsilon},
\label{eq:app:PV}
\end{equation}
\begin{equation}
s_-^\ell(\omega)=-\frac{\Gamma}{2\pi}{\cal P}\int_{-W}^{W}\dd\epsilon\,
\frac{f_{\mu_\ell}(\epsilon)}{\omega-\epsilon},
\label{eq:app:PVminus}
\end{equation}
where $f_\mu(\epsilon)=[\exp((\epsilon-\mu)/k_BT)+1]^{-1}$.
Writing $g^\pm(\delta,\mu)=-s_\pm^\ell(\delta)$ for $\mu=\mu_\ell$,
we obtain the wide-band expressions
\begin{equation}
\begin{aligned}
g^-(\delta,\mu)&\simeq\frac{\Gamma}{2\pi}
\biggl[\ln\left(\frac{W+\delta}{2\pi k_BT}\right)\\
&\quad-\operatorname{Re}\psi\left(\frac12+\ii
\frac{\delta-\mu}{2\pi k_BT}\right)\biggr],
\end{aligned}
\label{eq:app:digamma}
\end{equation}
\begin{equation}
g^+(\delta,\mu)=g^-(\delta,-\mu),
\label{eq:app:digammaplus}
\end{equation}
where $\psi$ is the digamma function. The wide-band approximation requires
chemical potentials and relevant detunings well away from the band edges. Equation~\eqref{eq:app:digammaplus} follows from
$\epsilon\to-\epsilon$ and $1-f_\mu(\epsilon)=f_{-\mu}(-\epsilon)$.

Independent reservoirs have no correlations between leads, so their
second-order Lamb shifts add:
\begin{equation}
H_{\rm LS}=\sum_{\ell=L,R}H_{\rm LS}^{\ell}(\mu_\ell),
\label{eq:app:HLS}
\end{equation}
where
\begin{equation}
H_{\rm LS}^{\ell}(\mu_\ell)
=\sum_{s,\eta,\omega}s_\eta^\ell(\omega)
J_\eta^{\ell s\dagger}(\omega)J_\eta^{\ell s}(\omega).
\label{eq:app:HLSlead}
\end{equation}
At fixed dot and tunneling parameters, $\mu_\ell$ enters only through the
Fermi functions in Eqs.~\eqref{eq:app:PV} and \eqref{eq:app:PVminus}.
Each contribution commutes with $H_{\rm dot}$ and retains the complete
virtual spectrum through the unrestricted operators in
Eqs.~\eqref{eq:app:eigenoperators} and \eqref{eq:app:eigenoperatorsminus}.
Projection onto the ground multiplet gives
\begin{equation}
PH_{\rm LS}^{\ell}(\mu_\ell)P
=h_{0,\ell}P+\bm h_\ell(\mu_\ell)\cdot\hat{\bm\tau}.
\label{eq:app:leadfield}
\end{equation}
The components are
\begin{equation}
h_{0,\ell}=\frac14\Tr_P\!\left[PH_{\rm LS}^{\ell}(\mu_\ell)P\right],
\end{equation}
\begin{equation}
\bm h_\ell(\mu_\ell)=\frac14\Tr_P\!\left[
\hat{\bm\tau}PH_{\rm LS}^{\ell}(\mu_\ell)P\right].
\end{equation}
Here $\Tr_P$ includes both orbital and spin states. The normalization
follows from $\Tr_P P=4$ and $\Tr_P(\hat\tau_a\hat\tau_b)=4\delta_{ab}$.
By linearity, $\bm h=\sum_\ell\bm h_\ell(\mu_\ell)$, as in
Eq.~\eqref{eq:lambfield}.

The channel restriction applies only to sequential tunneling involving
$P=\Pi_{E_0}$. Let ${\cal T}_P$ contain the energy pairs $(E_0,E')$ and
$(E',E_0)$ connecting $P$ to $Q_2,Q_4$. The restricted transition
operators are
\begin{equation}
\bar J_{\eta P}^{\ell s}(\omega)
=\sum_{\substack{(E,E')\in{\cal T}_P\\E-E'=\omega}}
\Pi_EJ_\eta^{\ell s}(\omega)\Pi_{E'}.
\label{eq:app:restrictedjumps}
\end{equation}
The associated jump operators are
\begin{equation}
L_\alpha=\sqrt{\gamma_\eta^\ell(\omega)}\bar J_{\eta P}^{\ell s}(\omega),
\end{equation}
where $\alpha=(\ell,s,\eta,\omega)$. Using
${\cal D}[L]X=LXL^\dagger-\{L^\dagger L,X\}/2$, we write
\begin{equation}
{\cal D}_P=\sum_\alpha{\cal D}[L_\alpha],
\end{equation}
\begin{equation}
{\cal D}_{\rm rest}=\sum_\beta{\cal D}[L_\beta],
\end{equation}
where $PL_\beta=L_\beta P=0$. The sequential-tunneling dissipation is
\begin{equation}
{\cal D}_{\rm seq}={\cal D}_P+{\cal D}_{\rm rest}.
\label{eq:app:dissipator}
\end{equation}
The jumps $L_\beta$ include their tunneling rates and retain all remaining
resolved sequential-tunneling channels. Transitions from $P$ to higher
charged multiplets are off shell in the bias window and are omitted, but their virtual contributions remain in
Eq.~\eqref{eq:app:HLSlead}. We assume that channels involving $P$ have
Bohr frequencies distinct from those of unrelated channels; accidental
coincidences would require joint jump operators.
The master equation on the full dot Fock space is then
\begin{equation}
\dot\varrho={\cal L}_{\rm seq}[\varrho]=\frac1{\hbar}\left[-\ii[H_{\rm dot}+H_{\rm LS},\varrho]
+{\cal D}_{\rm seq}(\varrho)\right].
\label{eq:app:GKSL}
\end{equation}
This GKSL form with nonnegative rates ensures completely positive,
trace-preserving dynamics, with $\varrho$ defined on the full dot Fock
space throughout.

\section{Bloch equation for the orbital pseudospin}
\label{app:bloch}

Projecting Eq.~\eqref{eq:app:GKSL} onto the ground multiplet gives the
evolution of $\rho_P=P\varrho P$. Since $PH_{\rm dot}P=E_0P$, the isolated
Hamiltonian does not contribute within this block. The remaining terms
describe Lamb-shift precession and tunneling into and out of $P$.
With $H_P=PH_{\rm LS}P$, we obtain
\begin{equation}
\dot\rho_P=-\frac{\ii}{\hbar}[H_P,\rho_P]
+\frac1{\hbar}\left({\cal G}_P[\varrho]-\frac12\{{\cal K}_P,\rho_P\}\right),
\label{eq:app:gainloss}
\end{equation}
where
\begin{equation}
{\cal G}_P[\varrho]=\sum_\alpha PL_\alpha Q_{\rm ch}\varrho
Q_{\rm ch}L_\alpha^\dagger P,
\label{eq:app:gainK}
\end{equation}
\begin{equation}
{\cal K}_P=\sum_\alpha PL_\alpha^\dagger Q_{\rm ch}L_\alpha P.
\end{equation}
The retained jumps connect $P$ only to $Q_{\rm ch}$, while
$P{\cal D}_{\rm rest}(X)P=0$. Thus ${\cal G}_P[\varrho]$ describes gain from
the charged multiplets, and ${\cal K}_P$ determines loss from $P$.

Define the ground-multiplet probability $w=\Tr\rho_P$ and, for $w>0$,
the conditional pseudospin $r_a=\Tr(\hat\tau_a\rho_P)/w$, where
$\hat\tau_a$ acts as a Pauli operator in orbital space. In the
spin-rotation-invariant setting considered here, the conditional state
factorizes as
\begin{equation*}
\frac{\rho_P}{w}=\rho_{\rm orb}\otimes\frac{\mathbb I_{\rm spin}}2,
\end{equation*}
where
\begin{equation*}
\rho_{\rm orb}=\frac12\left(\mathbb I_{\rm orb}
+\sum_a r_a\tau_a^{\rm orb}\right).
\end{equation*}
Using $P=\mathbb I_{\rm orb}\otimes\mathbb I_{\rm spin}$, we obtain
\begin{equation}
\rho_P=\frac w4(P+\bm r\cdot\hat{\bm\tau}).
\label{eq:app:block}
\end{equation}
The projected Lamb Hamiltonian can likewise be written as
\begin{equation}
H_P=h_0P+\bm h\cdot\hat{\bm\tau},
\end{equation}
with
\begin{equation}
h_a=\frac14\Tr(\hat\tau_aH_P).
\end{equation}
The normalization follows from $\Tr P=4$ and
$\Tr(\hat\tau_a\hat\tau_b)=4\delta_{ab}$.
Expanding the gain and loss operators in the same basis gives
\begin{equation}
{\cal G}_P[\varrho]=\frac14(g_0P+\bm g\cdot\hat{\bm\tau}),
\label{eq:app:gainlosscoefficients}
\end{equation}
\begin{equation}
{\cal K}_P=\gamma P+\bm\xi\cdot\hat{\bm\tau},
\end{equation}
where $g_0=\Tr{\cal G}_P[\varrho]$,
$g_a=\Tr(\hat\tau_a{\cal G}_P[\varrho])$,
$\gamma=\Tr {\cal K}_P/4$, and $\xi_a=\Tr(\hat\tau_a{\cal K}_P)/4$.
Positivity of ${\cal K}_P$ implies $\gamma\geq|\bm\xi|$.

Tracing Eq.~\eqref{eq:app:gainloss} with $\hat\tau_a$ and applying the
Pauli algebra yields
\begin{equation}
-\ii\Tr\bigl(\hat\tau_a[H_P,\rho_P]\bigr)
=2w(\bm h\times\bm r)_a
\end{equation}
and
\begin{equation}
\Tr\!\left[\hat\tau_a\left({\cal G}_P[\varrho]
-\frac12\{{\cal K}_P,\rho_P\}\right)\right]
=g_a-w\xi_a-w\gamma r_a.
\end{equation}
With the source $\bm\sigma(t)=\bm g[\varrho(t)]/w(t)-\bm\xi$, the
orbital moment obeys the Bloch equation
\begin{equation}
\frac{\dd(w\bm r)}{\dd t}
=\frac w{\hbar}\left[2\bm h\times\bm r
+\bm\sigma(t)-\gamma\bm r\right].
\label{eq:app:projectedmoment}
\end{equation}
Here $\bm h$ drives precession and $\gamma$ damps the moment. The source
$\bm\sigma$ combines gain, $\bm g/w$, with the orbital dependence of loss,
$-\bm\xi$.

\iffalse
Taking the trace without $\hat\tau_a$ gives
$\hbar\dot w=g_0-w(\gamma+\bm\xi\cdot\bm r)$.
Equation~\eqref{eq:app:projectedmoment} therefore evolves the moment $w\bm r$;
the equation for the conditional vector $\bm r$ contains the additional
term $-(\dot w/w)\bm r$.
\fi

At zero flux, tunneling pumps an in-plane orbital polarization, which
Lamb-shift precession converts into chirality. To see this, we first show
that $\sigma_z(t)=g_z[\varrho(t)]/w(t)-\xi_z=0$.

The loss operator ${\cal K}_P$ is real, whereas
$K\hat\tau_z K=-\hat\tau_z$, so $\xi_z=\Tr(\hat\tau_z{\cal K}_P)/4=0$.
For the gain term, cyclicity of the trace gives
\begin{equation}
g_z[\varrho]=\Tr(A_z\varrho),\qquad
A_z=\sum_\alpha Q_{\rm ch}L_\alpha^\dagger
P\hat\tau_zP L_\alpha Q_{\rm ch}.
\label{eq:app:Az}
\end{equation}
Since $A_z$ is a spin scalar and $Q_2$ and $Q_4$ each carry a
one-dimensional orbital representation, it has the form
$A_z=c_2Q_2+c_4Q_4$, with real coefficients. Both projectors are real,
so this expression is even under $K$. Yet the real jump operators and
the odd parity of $\hat\tau_z$ imply $KA_zK=-A_z$.
Hence $c_2=c_4=0$ and $g_z[\varrho]=0$ for any state of the charged
multiplets.

The energy projectors and tunneling operators are also real in the
occupation basis, making $H_P$ real. The same parity argument therefore
gives $h_z=0$. These results give
\begin{equation}
g_z[\varrho]=\xi_z=0,\qquad
\sigma_z(t)=h_z=0.
\label{eq:app:axialzero}
\end{equation}
The source thus acts within the pseudospin plane, while the chiral
moment evolves as
\begin{equation}
\hbar\frac{\dd(wr_z)}{\dd t}
=2w(h_xr_y-h_yr_x)-\gamma wr_z.
\label{eq:app:axialmoment}
\end{equation}
The first term is the Lamb torque: when the field and in-plane
polarization are not collinear, precession generates a chiral moment.
The second term describes its damping.

At stationarity, $\dot w=0$ and
$\bm\sigma(t)\to\bm\sigma_{\rm ss}$. Writing $\bm\sigma=\bm\sigma_{\rm ss}$,
Eq.~\eqref{eq:app:projectedmoment} reduces to
$\gamma\bm r_{\rm }=2\bm h\times\bm r_{\rm }+\bm\sigma$.
For $h_z=\sigma_z=0$, the components satisfy
\begin{equation}
\gamma r_x=\sigma_x+2h_yr_z,
\label{eq:app:stationarycomponents}
\end{equation}
\begin{equation}
\gamma r_y=\sigma_y-2h_xr_z,
\end{equation}
\begin{equation}
\gamma r_z=2(h_xr_y-h_yr_x),
\end{equation}
where $\bm r=\bm r_{\rm ss}$. Eliminating $r_x$ and $r_y$ gives
$(\gamma^2+4h_x^2+4h_y^2)r_z=2(h_x\sigma_y-h_y\sigma_x)$, hence
\begin{equation}
r_{z}=\frac{2(h_x\sigma_y-h_y\sigma_x)}
{\gamma^2+4|\bm h|^2}.
\label{eq:app:rz}
\end{equation}
The current--chirality relation in Eq.~\eqref{eq:app:exact} then gives
\begin{equation}
I_c^{(P)}=\lambda\chi_qwr_z.
\label{eq:app:Iprojected}
\end{equation}
The sign changes of $(\bm h\times\bm\sigma)_z$ therefore determine the
reversals of the projected current, while $w$ scales its amplitude.

For identical lead spectra and linewidths, reversing the bias is
equivalent to exchanging dots 2 and 3 through the reflection $\mathcal M$:
$\varrho_{\rm ss}(-V)=\mathcal M\varrho_{\rm ss}(V)\mathcal M^\dagger$.
Since $P\mathcal M P=\hat\tau_x$, this reflection preserves $r_x$ and
reverses $r_y$ and $r_z$, giving $\bm r\mapsto R\bm r$ with
$R=\operatorname{diag}(1,-1,-1)$. Thus $w(-V)=w(V)$ and
\begin{equation}
\bm r_{\rm ss}(-V)=R\bm r_{\rm ss}(V),
\label{eq:app:mirror}
\end{equation}
\begin{equation}
\bm h(-V)=R\bm h(V),
\end{equation}
\begin{equation}
\bm\sigma_{\rm ss}(-V)=R\bm\sigma_{\rm ss}(V).
\end{equation}
At zero bias, $h_y(0)=\sigma_y(0)=0$. Together with $h_z=\sigma_z=0$,
this makes the field and source collinear, so Eq.~\eqref{eq:app:rz} gives
$r_z(0)=0$. Equation~\eqref{eq:app:Iprojected} also yields
$I_c^{(P)}(-V)=-I_c^{(P)}(V)$.

\section{Gate dependence and validity of the ground-multiplet description}
\label{app:gate}

\begin{figure}[t]
\centering
\includegraphics[width=\columnwidth]{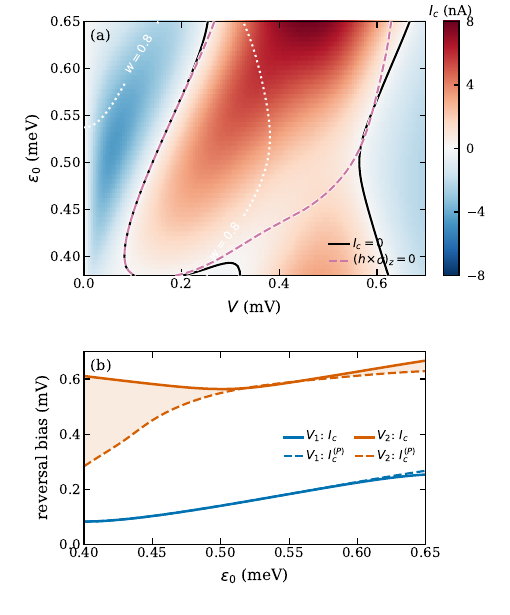}
\caption{Gate dependence of the circulating current and its reversals.
(a) Total circulating current $I_c$ as a function of bias and gate energy. Black solid contours indicate $I_c=0$, while magenta dashed contours identify the alignment condition $(\bm h\times\bm\sigma_{\rm ss})_z=0$. White dotted contours indicate a ground-multiplet population of $w=0.8$.
(b) Reversal biases $V_1$ and $V_2$ determined from the zero crossings of the total current $I_c$ (solid) and its ground-multiplet contribution $I_c^{(P)}$ (dashed).
\label{fig:app:gate}}
\end{figure}

To examine the gate dependence of current reversals, we evaluate the total
current $I_c$ and its ground-multiplet contribution $I_c^{(P)}$ from the
full stationary RDM at each gate energy and bias
[Eq.~\eqref{eq:app:currentdecomp}].

At low bias, the ground multiplet dominates the stationary population,
and the total-current reversal closely follows the alignment condition
$(\bm h\times\bm\sigma_{\rm ss})_z=0$ [Fig.~\ref{fig:app:gate}(a)].
The agreement between the low-bias reversal branches in
Fig.~\ref{fig:app:gate}(b) shows that dissipative pumping and Lamb-induced
precession account for the reversal across the gate range studied.

As the bias increases, the ground-multiplet population decreases and
other sectors contribute appreciably. The total current then vanishes
when
\begin{equation}
I_c^{(P)}=-I_c^{(\rm rest)},
\end{equation}
rather than when $I_c^{(P)}=0$, allowing the high-bias reversal branches
to separate. This separation reflects the growing contribution from
states outside the ground multiplet; the current--chirality relation
within $P$ remains valid.

\section{Comparison of master-equation and HEOM results}
\label{app:comparison}

\begin{figure}[t]
\centering
\includegraphics[width=\columnwidth]{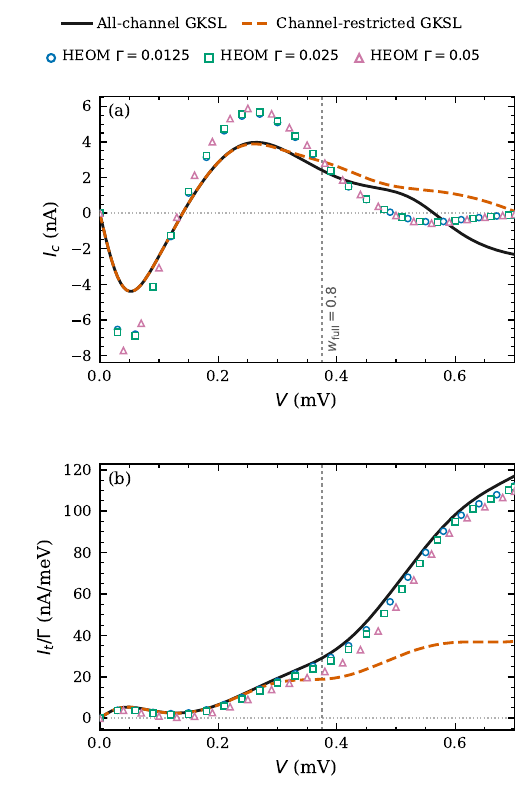}
\caption{Stationary currents obtained from second-order secular Born--Markov GKSL equations and HEOM.
(a) Total circulating current.
(b) Transport current normalized by $\Gamma$.
Black solid curves include all sequential-tunneling transitions among the two-, three-, and four-electron states. Orange dashed curves restrict transitions involving the ground multiplet $P$ to $P\leftrightarrow Q_2$ and $P\leftrightarrow Q_4$, while retaining all other transitions.
Open symbols show HEOM results for the indicated values of $\Gamma$.
The vertical dotted line indicates the bias at which the ground-multiplet population falls to $w=0.8$ in the calculation retaining all transitions.
The parameters are the same as in Fig.~\ref{fig:benchmark}.
\label{fig:app:generator}}
\end{figure}

Figure~\ref{fig:app:generator} compares stationary currents from HEOM and
two second-order secular Born--Markov master equations. One includes all
sequential-tunneling transitions among the two-, three-, and four-electron
states. The other restricts transitions involving the ground multiplet
$P$ to $P\leftrightarrow Q_2$ and $P\leftrightarrow Q_4$, while retaining
all other transitions. Both preserve coherences within degenerate
energy subspaces and include the same complete virtual spectrum in the
Lamb shift. In each calculation, the circulating current is obtained from
the full stationary RDM as $I_c=\Tr(\varrho_{\rm ss}\hat I_c)$.

At low bias, the two master equations yield nearly identical circulating
currents and reproduce the reversal found in HEOM
[Fig.~\ref{fig:app:generator}(a)], although their amplitudes differ from
the HEOM values. At higher bias, the curves separate as additional
sequential-tunneling channels contribute. Including these channels changes
the circulating current but does not remove the quantitative differences
from HEOM.

For the transport current [Fig.~\ref{fig:app:generator}(b)], the HEOM
curves nearly coincide after normalization by $\Gamma$, consistent with
a leading contribution linear in the linewidth. The two master equations
agree at low bias, whereas restricting the transitions involving $P$
underestimates the high-bias current. Including all sequential-tunneling
transitions substantially improves agreement with HEOM in this regime.

The difference between the two master equations isolates the effect of
the omitted sequential-tunneling channels. Residual differences from HEOM
may also reflect the second-order, Markov, and secular approximations,
through higher-order tunneling, reservoir memory, and neglected
coherences. These contributions cannot be distinguished by the present
comparison. The low-bias circulating-current reversal is nevertheless
common to all three calculations.

\bibliography{references}

\end{document}